\documentclass{jltp}

\title{Testing the Zurek-Kibble Causality Bounds with Annular Josephson Tunnel
Junctions}

\author{Roberto Monaco, Ray Rivers$^*$ and Eleftheria Kavoussanaki$^*$}

\address{ Istituto di Cibernetica del C.N.R., I-80072, Arco Felice (Na), Italy%
\\
and INFM-Dipartimento di Fisica, Universita' di Salerno, I-84081 Baronissi\\
(Sa), Italy\\
$^*$Theoretical Physics, Blackett Laboratory, Imperial College, Prince\\
Consort Road, London, SW7 2BZ, U.K.}

\runninghead{\bfseries R.~Monaco, R.J.~Rivers and E.~Kavoussanaki}
{Testing the ZK causality bounds with annular JTJs}

\begin{document}

\maketitle

\begin{abstract}
Zurek has provided simple causal bounds for the onset of phase transitions
in condensed matter, that mirror those proposed by Kibble for relativistic
quantum field theory. In this paper we show how earlier experiments with annular
Josephson tunnel Junctions are consistent with this scenario, and suggest how
further experiments might confirm it.

PACS numbers: 11.27.+d, 05.70.Fh, 11.10.Wx, 67.40.Vs
\end{abstract}

\section{INTRODUCTION}

As the early universe went through a period of rapid cooling it
changed phase several times. A signal of such transitions would be
the direct or indirect observation of topological defects (e.g.
magnetic monopoles) arising from the inhomogeneities of the
ordered phase. Such defects appear because the correlation lengths
$\xi (t)$ of the order parameter fields are necessarily {\it
finite} for transitions implemented in {\it finite} time. This is
even though, for the continuous transitions that we have in  mind,
the equilibrium correlation lengths $\xi_{eq}$ would diverge at
the critical temperature if there were time enough (the adiabatic
approximation).

Kibble \cite{kibble1,kibble2} observed that this simple causality
imposed useful constraints on domain growth and the density of
defects at the time of their formation. Unfortunately, because of
our lack of knowledge about the details of the transitions in the
early universe it is impossible to find reliable predictions for
current observations. However, since notions of causality are not
specific to the relativistic quantum field theory (QFT)
appropriate to the early universe, Zurek \cite{zurek1,zurek2}
argued that similar causal bounds were valid in condensed matter
systems for which direct experiments on defect formation could be
performed. These bounds can be understood in several ways. For our
purposes the most useful is the following:

As an ordering transition begins to be implemented, from time
$t=0$, say, there is a maximum speed $c(t)$ at which the system
can become ordered. This leads to causal horizons, outside of
which there are no correlations. Zurek proposed that the earliest
time (the 'causal time') at which defects can form is when a {\it
local} causal horizon is large enough to accommodate a single
defect at that time. Since the defect size $\xi (t) =\xi (T(t))$
depends on temperature $T$, which depends on time, the causal time
${\bar t}$ is determined from
\begin{equation}
 \xi ({\bar t})\approx 2\int^{\bar t}_0 dt\,c(t). \label{xic}
\end{equation}

 This time is not to be confused with the earlier time,
that
 depends on the {\it global} structure of the system,
 when the coherence length becomes smaller than the system.
 The two successful experiments\cite{helsinki,grenoble} on defect
 production in superfluid $^3 He$
 give very clear evidence for the primacy of the causal horizon in
 determining when defects form.

There are other ways to formulate causality\cite{zurek2}, most
simply by requiring that $\xi (t)$ cannot grow faster than $c(t)$.
This can be imposed either before the transition, or after. In
simple systems all bounds agree, up to numerical factors
approximately unity\cite{zurek2}. To the level at which we are
working (better than an order of magnitude, but to a factor of a
few), they are indistinguishable.

Whatever the case, although the causal bounds are robust, the
extent to which they are saturated depends on the details of the
microscopic dynamics. From the microscopic level, causality along
the lines above is not explicit, although encoded in the relevant
dynamical equations. The picture is rather one of order being
established through the growth of the amplitudes of
long-wavelength instabilities. The earliest time at which we can
identify defects from this viewpoint is when the order parameters
have achieved their equilibrium magnitudes. Qualitatively, for
simple models this time is also in agreement with the causal time
above, showing that the causal bounds are approximately saturated.
There is no real surprise in this. The causal time  and distance
scales ${\bar t}$ and ${\bar\xi} = \xi({\bar t})$ are just as we
would expect from dimensional analysis (in the mean-field
approximation) and unstable modes grow exponentially, whereby the
dependence of the causal time (and corresponding defect density)
on the microscopic parameters is only (square-root) logarithmic.
However, before the causal time we now have a picture in which
there is a fractal thermal fuzz of potential defects, whose
density depends on the scale at which we look. By the causal time
some of these have matured into the (scale-independent) defects
that we see subsequently. This is seen in numerical
simulations\cite{iba}, and see the article by one of us (RR) in
these proceedings\cite{rivers2}, where this is discussed at
greater length.

The experimental confirmation\cite{helsinki,grenoble} of the
Kibble-Zurek predictions for vortex production in superfluid
$^{3}$He, at better than order-of-magnitude level, reinforced the
hope that, if we understood why the bounds were approximately
respected there and elsewhere, we might have a better
understanding of whether they would be respected in QFT. While
$^{3}$He has an unparalleled richness in its order parameters
\cite{grisha} and the topological defects that can be produced in
it, it is a difficult system in which to understand the
fundamentals of defect formation. As an independent test of his
assumptions Zurek \cite{zurek2} also proposed experiments to
measure spontaneous superflow in $^{4}$He and spontaneous flux
generation in superconductors (essentially topological charge
density) in much simpler one-dimensional annular geometries, for
which predictions are easier to make. In practice, correlation
lengths in both these systems are so small at the relevant
timescales that it has been difficult to make reasonably
1-dimensional systems, and the experiments have not been
performed.

It is with this in mind that, in this paper, we shall examine the
spontaneous generations of defects (`fluxons') in annular
Josephson tunnel Junctions (JTJs) for which correlations lengths
are sufficiently large that effective 1-dimensional systems can be
fabricated and the Zurek predictions can be tested. One of us (RM)
has conducted such experiments \cite{Roberto} in JTJs in the
context of a different experimental programme that was completed
before the suggestion was made that the Zurek bounds were relevant
in that case.

 Although
the choice of materials was not optimised for the Zurek scenario
the results were very striking. Firstly, defects were, indeed,
detected on quenching the system. This is in contrast to their
non-observance in some other experiments, on planar
superconductors\cite{Polturak} and\cite{lancaster2} $^4 He$,
specifically designed to check the
Zurek picture.
 Secondly, although the circumference of the ring
was at least three orders of magnitude larger than a fluxon, their
production was rare. A single fluxon only appeared a few percent
of the time, two very rarely, and more than two, never.  At the
time there was no understanding as to why it was so difficult to
produce fluxons and no way to estimate the likelihood of their
appearance. For the purpose of the experiments this did not
matter, since those that were produced were then used for other
purposes\cite{Roberto}. We aim to show that these experimental
results are compatible with Zurek's predictions which, to date,
provide their only quantitative explanation. However, because of
the compound nature of JTJs, causal bounds and their saturation is
more complicated.

The possibility that the Zurek-Kibble analysis could be applied to
JTJs was first considered briefly\cite{RKK} by two of us (EK and
RR), and later a more substantial analysis was reported elsewhere
\cite{KMR}. This paper provides an extended and updated
presentation of the theory and future experiments.

\section{JOSEPHSON TUNNEL JUNCTIONS}

An annular JTJ consists of two superimposed annuli of ordinary
superconductors of thickness d$_{s}$, width $\Delta r$, separated
by a layer of oxide of thickness d$_{ox}$, with relative
dielectric constant $\epsilon _{r}$. We consider a quench in which
the (spatially uniform) temperature $T(t)$ of the JTJ varies in
time as
\begin{equation}
\frac{dT}{dt}=-\frac{T_{c}}{\tau _{Q}} \label{dT}
\end{equation}
in the vicinity of $T_{c}=T(0)$, $\tau _{Q}$ being the quenching
time.

Above the transition, at temperature $T_c$, we have two normal
conductors but, after the transition has been completed, the
complex order parameters $\Psi _{1}=\rho _{1}e^{i\varphi _{1}}$
and $\Psi _{2}=\rho _{2}e^{i\varphi _{2}}$ of the individual
superconductors are characterised by the phases $\varphi _{1}$ and
$\varphi _{2}$ alone. The Josephson current density at any point
$x$ in the annulus is
\begin{equation}
J=J_{c}(T)\sin \varphi (x),
\label{J}
\end{equation}
where $\varphi (x)=\varphi _{1}(x)-\varphi _{2}(x)$, and
$J_{c}(T)$ depends on the nature of the junction. All other things
being constant, the larger the resistance of the oxide, the weaker
the coupling between the superconductors and the smaller the
critical current $J_c(0)$.

\subsection{The individual superconductors}
Before considering the effect of a quench on a JTJ it is helpful,
for purposes of comparison, to consider the effect of the same
quench (\ref{dT}) on each of the individual superconductors, in
the limit (vanishing critical current $J_c$) in which they are not
connected. We also take the limit in which their thickness is
smaller than any other scale.

This has been analysed already\cite{zurek2} by Zurek, and we
repeat his conclusions.  Consider a single superconducting ring
with scalar order parameter $\Psi =\rho e^{i\varphi}$. We are
being simplistic here in our evasion of the gauge properties of
the theory\cite{Mark}. However, for strong second-order
transitions it should be true - essentially we are looking for a
benchmark. In a mean-field approximation the relevant adiabatic
correlation length has the behaviour\cite{zurek2}
\begin{equation}
\xi (t)=\xi (T(t))=\xi _{0}\sqrt{\frac{\tau _{Q}}{t}} \label{xit}
\end{equation}
in terms of the fundamental length scale $\xi_0$ (the size of a
cold defect, among other things). The maximum speed at which the
field can order itself is
\begin{equation}
c(t)=c(T(t))=C_{0}\sqrt{\frac{t}{\tau _{Q}}%
}, \label{ct}
\end{equation}
where $C_0=\xi_0/\tau_0$ is given in terms of $\tau_0$, the
relaxation time for long wavelength modes. [We save the symbol
$c_0$ for the speed of light in vacuo.] Note that $c(t)$ shows
critical slowing down.

Imposing the causal bound (\ref{xic}) gives a causal time ${\bar
t}_s$ for individual superconductors of, approximately,
\begin{equation}
{\bar t}_s = \sqrt{\tau_0\tau_Q}, \label{ts}
\end{equation}
at which time the causal horizon is
\begin{equation}
\overline{\xi }_{s}=\xi(\overline{t}_s)=\xi _{0}\sqrt{\frac{\tau
_{Q}}{\overline{t}_s}}=\xi _{0}(\frac{\tau _{Q}}{\tau
_{0}})^{\frac{1}{4}}. \label{xibars}
\end{equation}

At these very early times the single length $\overline{\xi }_{s}$
serves to set both the coherence length in $\rho$, as it achieves
its equilibrium value, and in $\varphi$. With each change in
$\varphi$ of $2\pi$ will be associated with a unit of flux within
the annulus. At the risk of causing some confusion, we could also
term these fluxons. On average, the number $N$ of such fluxons
will be zero. However, the variance in $N$ will be non-zero. If,
as a first guess, we assume that the phase $\varphi$ takes a
random walk around the annulus, circumference $C = 2\pi{\bar r}$,
then
 for $C\gg
{\bar\xi}_s$,
\begin{equation}
\Delta N =\frac{\Delta \varphi }{2\pi }= O\bigg( \frac{1}{2\pi
}\sqrt{\frac{C} {{\bar\xi}_{s}}}\bigg). \label{Dn0}
\end{equation}
There are many qualifications to (\ref{Dn0}), but this is a good
starting point.

Numerically, Zurek proposed a system-independent estimate for
$\tau_0$ from Gorkov's equation as
$\tau_{0}=\pi \hbar /16k_{B}T_{c}\approx 0.15\thinspace$ps for $%
T_{c}\simeq$ 10\thinspace K. The end result is that, for $\tau_Q$
in  seconds and $T_c$ in degrees Kelvin,
\begin{equation}
{\bar t}_s \approx \sqrt{\tau_Q/T_c}\, [\mu\thinspace\mbox{s}].
\label{tbars}
\end{equation}
For $\tau_Q = 1\thinspace$s and $T_c = 10$\thinspace K, ${\bar
t}_s$ is a fraction of a microsecond.
 Further, with a reasonable choice of $\xi_0\approx
100$\thinspace nm, we have ${\bar\xi}_s\approx 0.1$\thinspace mm.

When we first tried to understand whether the Zurek-Kibble bounds
had any relevance to the experiments of
Ref.\thinspace\onlinecite{Roberto}, two of us\cite{RKK} made the
naive approximation that, up to the formation of fluxons, we could
treat the superconductors as independent, with uncorrelated phase
angles $\varphi_1$ and $\varphi_2$. In that case,
$(\Delta\varphi)^2 =(\Delta\varphi_1)^2 + (\Delta\varphi_2)^2$.
For a ring of circumference $0.5\thinspace$mm, as in the
experiments, we would predict $\Delta N = O(1)$ from (\ref{Dn0}).
This is an order of magnitude too large, but there is a certain
amount of latitude in all these estimates, and it did not seem
totally unreasonable.

However, what made the approximation untenable was that, on
keeping the superconductors otherwise identical, but changing the
coupling between them, no fluxons were seen at a level at which
they would have been seen, had the superconductors behaved
independently. We therefore have to turn to coupled junctions,
where we find a reason for this.

\subsection{Josephson tunnel junctions}

In order to establish causality bounds for JTJs we need to
identify the velocity that establishes the size of the causal
horizons and the length that characterises the fluxon size.

The strength of Zurek's predictions lies in the assumption that,
in general, these can be read off from the time and space
derivatives {\it alone} in the equations of motion in the
adiabatic regime. To see how well the bounds are saturated
requires further knowledge, but we shall assume that they will be
well saturated here, as exemplified by the $^3He$ experiments.
Further, in this scenario the adiabatic approximation is pushed to
its limits, by which we take it to be approximately valid from the
causal time onwards.

 If we ignore dissipative effects from quasiparticle tunneling and surface losses
 in the adiabatic regime at temperature $T<T_c$, $\varphi $
satisfies the
 {\it one-dimensional} Sine-Gordon (SG) equation\cite{Lomdahl}
\begin{equation}
\frac{\partial ^{2}\varphi }{\partial x^{2}}-\frac{1}{\overline{c}^{2}(T)}%
\frac{\partial ^{2}\varphi }{\partial t^{2}}
=\frac{1}{\lambda _{J}^{2}(T)}\sin
\varphi,
\label{SG}
\end{equation}
 provided the width
$\Delta r$ of the annulus, of radius ${\bar r}$, satisfies $\Delta
r\ll{\bar r} $ and $\Delta r\ll\lambda _{J}(T)$, the Josephson
coherence length. In this case $x$ measures the distance along the
annulus, with periodic boundary conditions. Unlike the case of
superfluid $^{4}$He and annular (single) superconductors,
it is not difficult to implement $%
\Delta r\ll\lambda _{J}(T)$, as we shall see.

As with other established models of defect formation, the
classical equations  are only valid once the transition is
complete. We shall not use the SG equation, together with its
dissipative and other terms, to study the evolution of defects. In
fact, we do not study their evolution at all. However,
 Eqs.\thinspace\ref{SG} and \ref{fluxon} are sufficient to enable us, in
the spirit of the Zurek-Kibble scenario, to identify $\lambda_J
(T)$, diverging at $T_c$, as the equilibrium correlation length
$\xi_{eq}$ to be constrained by causality.  Further, the Swihart
velocity $\overline{c}(T)$ (with critical slowing down at
$T=T_c$), measures the maximum speed at which the order parameter
can change\cite{Swihart,Barone}. We shall see later that,
formally, $\overline{c}(t)$ has a behaviour similar to $c(t)$,
\begin{equation}
\overline{c}(t)=\overline{c}(T(t))=\overline{c}_{0}\sqrt{\frac{t}{\tau _{Q}}%
}. \label{cbar2}
\end{equation}
Away from the critical
temperature dissipative effects are small.

The classical topological defects (the JTJ fluxons) are the
'kinks' of the Sine-Gordon theory
\begin{equation}
\varphi _{\pm }(x,t)=4\arctan \exp \bigg[\pm \frac{x-ut}{\lambda _{J}(T)\sqrt{%
1-(u/{\bar c}(T))^{2}}}\bigg], \label{fluxon}
\end{equation}
where $|u|<\overline{c}(T)$.

In fact, (\ref{fluxon}) is oversimplified on many counts. For
example, the maximum speed of the JTJ fluxon is reduced in the
presence of an external magnetic field (which reduces its
behaviour to that of a damped pendulum). Nonetheless, at our level
of approximation, the Swihart velocity remains a good candidate
for determining the causal horizon, after very early times.

At very early times after the beginning of the transition ($t>0$),
before the individual superconductors have the ability to
superconduct, the causal bound is, most plausibly, driven
dissipatively by $c(t)$ of (\ref{ct}). If it were the case that
$C_0$ and $\overline{c}_{0}$ were equal, we would be happy to use
the formula (\ref{cbar2}) all the way down to $t=0$. Although
$\overline{c}_{0}$ depends on the nature of the junction, rather
than just the individual superconductors, for the junctions of
Ref.\thinspace\onlinecite{Roberto}, $\overline{c}_{0}\approx
10^7\thinspace$ m/s for the samples used. This is an order of
magnitude larger than $C_0$, as diagnosed earlier. However, we
note that the size of the causal horizon at the causal time only
depends on the scale of the velocity to its quarter power.
 To the level at which
we are working (better than an order of magnitude, but to a factor
of a few), we can take ${\bar c}(t)$ as the velocity in
(\ref{xic}).

Our horizon bound for the earliest time ${\bar t}$ that we can see
classical JTJ fluxons is when the causal horizon is big enough to
accommodate one. That is,
\begin{equation}
 \lambda_J ({\bar t})\approx 2\int^{\bar t}_0 dt\,{\bar c}(t)
 \label{lbar}
\end{equation}
at ${\bar t}={\bar t}_J$. This is consistent in that, at time
${\bar t}_J$, classical JTJ fluxons that have been formed will be
contracting at the Swihart velocity\cite{KMR}. To have attempted
to produce them before (as would happen with a slower velocity)
would make them want to contract faster than causality permits.

We shall give details later, but the end result is that ${\bar
t}_J\geq {\bar t}_s$, at our level of approximation. We now
understand better why our earlier assumption of independent
superconductors is not correct. As the superconductors establish
themselves, thermal fluctuations lead to the production of phase
rotations in each superconductor (as discussed in Section 2.1).
However, with the conductors connected, there will be continuous
cancellation between them. Flux lines of opposite direction can
annihilate in the oxide and disappear from the system to leave
only the residual locally unpaired flux lines, now classical JTJ
fluxons with correlation length ${\bar\lambda}_J$. They will be
reduced in number to our earlier guess, in agreement with
experiment.

\section{SYMMETRIC JTJs}

After these generalities, we need a specific model. For simplicity
we take a {\it symmetric} JTJ, in which the electrodes are made of
identical materials with common critical temperatures $T_{c}$ and
assume a quench in which the (spatially uniform) temperature
$T(t)$ of the JTJ varies in time as in (\ref{dT}).

We reiterate that, in our approximation, the Zurek-Kibble bounds
rely on the fact that the Swihart velocity is, numerically at
least, the maximum speed at which the order parameter can change
at any time. For speeds lower than this, we retain the
adiabatic approximation, in which the correlation length of the field $%
\lambda _{J}(t)$ is just $\lambda _{J}(T(t))$, where $\lambda _{J}(T)$ is
the adiabatic (equilibrium) coherence length at temperature T.

For a symmetric JTJ
\begin{equation}
\lambda _{J}(T)=\sqrt{\frac{\hbar }{2e\mu _{0}d_{e}(T)J_{c}(T)}},
\label{lJ}
\end{equation}
where $J_{c}(T)$ is the critical Josephson current density introduced
earlier and d$_{e}(T)$ is the magnetic thickness.\ As for the latter, if $%
\lambda _{L}(T)$ is the London penetration depth of the two (identical)
superconducting sheets,

\begin{equation}
\lambda _{L}(T)=\frac{\lambda _{L}(0)}{\sqrt{1-(\frac{T}{T_{c}})^{4}}},
\label{lL}
\end{equation}
then

\begin{equation}
d_{e}(T)=d_{ox}+2\lambda _{L}(T)\tanh \frac{d_{s}}{2\lambda _{L}(T)}.
\label{de}
\end{equation}
Neglecting the barrier thickness $d_{ox}\ll d_{s}$, $\lambda _{L}$, gives $%
d_{e}=d_{s}$ close to $T_{c}$, i.e the magnetic thickness equals the film
thickness and can be set constant in (\ref{lJ}).

All the T-dependence (and t-dependence) of $\lambda _{J}$ resides
in $J_{c}$, which has the form\cite{Barone}

\begin{equation}
J_{c}(T)=\frac{\pi }{2}\frac{\Delta (T)}{e\rho _{N}}\tanh \frac{\Delta (T)}{%
2k_{B}T}.
\label{Jc}
\end{equation}
In (\ref{Jc}), $\Delta (T)$ in the superconducting gap energy and varies steeply
near $T_{c}$ as

\begin{equation}
\Delta (T)\simeq 1.8\Delta (0)\sqrt{1-\frac{T}{T_{c}}}
\label{D}
\end{equation}
and $\rho _{N}$ is the JTJ normal resistance per unit area.

Introducing the dimensionless quantity $\alpha =1.6\Delta (0)/k_{B}T_{c}$,
whose typical value is between $3$ and $5$, enables us to write $J_{c}(T)$ as

\begin{equation}
J_{c}(T)\simeq \alpha J_{c}(0)(1-\frac{T}{T_{c}}).
\label{Jc2}
\end{equation}
Thus, on repeated substitution of $T(t)$ of (\ref{dT}) in the above, we see that in
the vicinity of the transition,

\begin{equation}
\lambda _{J}(t)=\lambda _{J}(T(t))={\bar\xi} _{0}\sqrt{\frac{\tau
_{Q}}{t}} \label{lJt}
\end{equation}
where

\begin{equation}
{\bar\xi} _{0}=\sqrt{\frac{\hbar }{2\varepsilon \mu
_{0}d_{s}\alpha J_{c}(0)}}. \label{xi}
\end{equation}
At the same time, for a finite electrode thickness JTJ, the Swihart velocity
takes the form

\begin{equation}
\overline{c}(T)=c_{0}\sqrt{\frac{d_{ox}}{\epsilon _{r}d_{i}(T)}},
\label{cbar}
\end{equation}
where

\begin{equation}
d_{i}(T)=d_{ox}+2\lambda _{L}(T)\coth (\frac{d_{s}}{2\lambda _{L}(T)}).
\label{di}
\end{equation}
In the vicinity of the transition it follows, for the quench
(\ref{dT}), that we recover $\overline{c}(t)$ of (\ref{cbar2}),
$\overline{c}(t)= {\bar c}_0\sqrt{t/\tau_Q}$, where
$\overline{c}_{0}=c_{0}\sqrt{d_{s}d_{ox}/\epsilon _{r}\lambda
_{L}^{2}(0)}$. Inspection of (\ref{lbar}) shows that
\begin{equation}
\overline{t}_J=\sqrt{{\bar\tau} _{0}\tau _{Q}} \label{tbar2}
\end{equation}
with ${\bar\tau} _{0}={\bar\xi} _{0}/{\bar c}_{0}$, in analogy
with (\ref{ts}).

Making the second assumption of Zurek and Kibble that $\lambda _{J}(%
\overline{t}_J)$ also characterises kink separation at this
formation time, we find
\begin{equation}
\overline{\lambda }_{J}=\lambda _{J}(\overline{t}_J)={\bar\xi}
_{0}\sqrt{\frac{\tau _{Q}}{\overline{t}_J}}={\bar\xi}
_{0}(\frac{\tau _{Q}}{{\bar\tau} _{0}})^{\frac{1}{4}},
\label{lbarJ}
\end{equation}
{\it identical} in form to (\ref{xibars}) for superconductors. The
quarter-power dependence of ${\bar\lambda}_J$ on ${\bar c}_0$ (for
given ${\bar\xi}_0$) that we alluded to earlier is apparent.
Moreover, if we keep ${\bar c}_0$ fixed, then
${\bar\lambda}_J\propto {\bar\xi}_0^{3/4}$.
Inserting reasonable values of ${\bar\xi} _{0}\geq 10\thinspace\mu$m and $\overline{c}%
_{0}=10^{7}\thinspace$m/s gives ${\bar\tau}_{0}\geq 1\thinspace$ps.
Thus, if $\tau _{Q}=$1\thinspace s we have $%
\overline{t}_J=1\thinspace\mu$s.
We see that $%
\overline{t}_{s}<\overline{t}_J$ by up to a factor of a few.

Whatever, we have a practical problem with symmetric JTJs in this
parameter range, in that substituting these values in
${\bar\lambda}_J$ of (\ref{lbarJ}) gives
 ${\bar\lambda}_J\approx 10$\thinspace mm. This  ${\bar\lambda}_J$
 which, in Zurek's picture, should characterise fluxon separation at the quench, would require
 very large annuli in order to see fluxons.

Fortunately, such concerns are theoretical since the fabrication of JTJs typically
yields {\it non-symmetric} devices with more acceptable properties.

\section{NON-SYMMETRIC JTJs}

Suppose the two superconductors, 1 and 2, now have different critical
temperatures $T_{c_{2}}>T_{c_{1}}$. Fluxons only appear at temperatures $%
T<T_{c_{1}}$, from which we measure our time $t$. At this time the
individual superconducting gap energies are

\begin{equation}
\Delta _{2}(T_{c_{1}})\simeq 1.8\Delta _{2}(0)\sqrt{1-\frac{T_{c_{1}}}{%
T_{c_{2}}}}
\label{D2}
\end{equation}
and $\Delta _{1}(T)$ vanishes at $t=0$ as

\begin{equation}
\Delta _{1}(T_{c_{1}})\simeq 1.8\Delta _{1}(0)\sqrt{\frac{t}{\tau _{Q}}}.
\label{D1}
\end{equation}
The critical Josephson current density $J_{c}^{\prime }(T(t))$ for a
non-symmetric JTJ is  now%
\cite{Barone}
\begin{equation}
J_{c}^{\prime }(T)=\frac{\pi \Delta _{1}(T)\Delta _{2}(T)}{\beta
(T)e\rho _{N}}\sum_{l=-\infty }^{\infty }\left\{ \left[ \omega
_{l}^{2}+\Delta _{1}^{2}(T)\right] \left[ \omega _{l}^{2}+\Delta
_{2}^{2}(T)\right] \right\} ^{-1/2},  \label{Jc2a}
\label{JprimeT0}
\end{equation}
\noindent where $\beta =1/k_{B}T$ and the $\omega _{l}=(2l+1)\pi
/\beta $ are the fermionic Matsubara frequencies. Near $t=0$
($T=T_{c,1}$)
\begin{equation}
J_{c}^{\prime }(T(t))\simeq \alpha
^{\prime }J_{c}^{\prime }(0)\sqrt{1-\frac{T_{c_{1}}}{T_{c_{2}}}}\sqrt{\frac{t}{\tau _{Q}}}
\label{JprimeT}
\end{equation}
instead of (\ref{Jc2a}), where
\begin{equation}
J_{c}^{\prime }(0)=\frac{\pi \Delta _{1}(0)\Delta
_{2}(0)}{[\Delta _{1}(0)+\Delta _{2}(0)]e\rho _{N}},\,\,\,\,\,\,\,\alpha ^{\prime }=%
\frac{[\Delta _{1}(0)+\Delta _{2}(0)]}{k_B T_{c_{1}}   },
\label{Jprime0}
\end{equation}
 provided $\Delta _{2}(T_{c_{1}})\ll 2\pi k_{B}T_{c_{1}}$.
 This is the case here.

If we now construct $\lambda _{J}(T(t))$ we find

\begin{equation}
\lambda _{J}(t)=\lambda _{J}(T(t))\approx {\bar\xi} _{0}(1-\frac{T_{c_{1}}}{%
T_{c_{2}}})^{-1/4}(\frac{\tau _{Q}}{t})^{1/4},
\label{lJt2}
\end{equation}
where ${\bar\xi} _{0}$ is as in (\ref{xi}), and we have used the fact that $%
J_{c}^{\prime }(0)\simeq J_{c}(0)$ and $\alpha ^{\prime }\simeq \alpha $.
The crucial difference between $\lambda _{J}(t)$ of (\ref{lJt2}) and $\lambda _{J}(t)
$ of (\ref{lJt}) for the symmetric case lies in the different critical index.
$\lambda _{J}(t)$ of (\ref{lJt2}) is very insensitive to the time at which
fluxons form.  For
the critical behaviour of (\ref{JprimeT}) to be valid, our earlier condition now
becomes $1-T_{c_{1}}/T_{c_{2}}\gg O(\overline{t}_J/\tau _{Q}%
)=O(10^{-6})$, which is always the case. The critical time $\overline{t}_J$,
determined from (\ref{tbar2}) as before, now gives

\begin{equation}
\overline{t}_J={\bar\tau} _{0}^{4/7}\tau _{Q}^{3/7}(1-\frac{T_{c_{1}}}{T_{c_{2}}}%
)^{-1/7}
\label{tbar3}
\end{equation}
rather than (\ref{tbar2}). For a typical value
$(1-T_{c_{1}}/T_{c_{2}})=0.02$ and the same values of ${\bar\tau}
_{0}$ and $\tau _{Q}$ as used previously, we find
$\overline{t}_J\approx 0.24\thinspace\mu$s. In turn, this gives

\begin{equation}
\overline{\lambda }_{J}=\lambda _{J}(\overline{t}_J)\simeq {\bar\xi} _{0}(1-\frac{%
T_{c_{1}}}{T_{c_{2}}})^{-1/4}(\frac{\tau _{Q}}{{\bar\tau}
_{0}})^{1/7}\simeq 1.4\thinspace\mbox{mm}. \label{lJbar}
\end{equation}
This is an order of magnitude {\it smaller} than in the symmetric
case, but still an order of magnitude larger than the
corresponding lengths in annuli of single superconductors.

However, we note that $\overline{t}_J$ of (\ref{tbar3}) is now slightly smaller than $\overline{t%
}_{s}$ of (\ref{tbars}). For all our caveats about the order of
magnitude
nature of the bounds, if we replace $\overline{t}_J$ by $\overline{t}%
_{s}$, $\overline{\lambda }_{J}$ of (\ref{lJbar}) is reduced
slightly to $1.1\thinspace$mm. The difference is negligible, given
the crudity of the bounds.

In summary, for annular JTJs of this type  we expect
$\overline{\lambda }_{J}\sim 1\thinspace$mm, rather than
$10\thinspace$mm or $100\thinspace\mu$m, and we would expect to
start to see fluxons once the circumference $C\sim 1\thinspace$mm.
This is, indeed, what happens.

\section{PAST AND FUTURE EXPERIMENTS}

As we said earlier, the original experiments\cite{Roberto}
performed by one of us (RM) were devised with other aims in mind.
The intention was to produce fluxons for further experiments and
the density and frequency with which they were produced was
secondary. The experiments that have been performed, and which we
have in mind, are simple in principle. As we cool an annular JTJ
uniformly from above its critical temperature $T_{c}$, domains in
$\varphi $ will form  whose boundaries are the fluxons. The mean
flux is zero.

The relevant observational quantity here is the variance in the
flux or, equivalently, the variance $\Delta N$ in the number of
fluxons {\it minus} antifluxons. This stability gives us a
distinct advantage over experiments in vortex production in
superfluid $^{4}$He  in which what is measured is a (decaying)
density of vortices {\it plus}
antivortices\cite{lancaster,lancaster2}. In fact the null result
of Ref.\onlinecite{lancaster2} for finding vortices in $^{4}$He
can be attributed to uncertainty over the decay rate of vortex
tangles\cite{rivers}.

If the circumference of the annulus is $C=2\pi{\bar r}$, and the
mean $\varphi $ correlation length at the time $\overline{t}_J$ of
the fluxon formation is $ {\bar\lambda}_{J}$ then, in parallel
with our earlier observations, we would expect, for $C\gg
{\bar\lambda}_J$,
\begin{equation}
\Delta N=\frac{\Delta \varphi }{2\pi }= O\bigg( \frac{1}{2\pi
}\sqrt{\frac{C} {{\bar\lambda}_{J}}}\bigg), \label{Dn}
\end{equation}
as a result of a random walk in phase around a ring. It is the
prediction (\ref{Dn}) that we would like to test for a variety of
junctions and temperature quenches, with different
$\overline{\lambda} _{J}$. Unfortunately, for the experiments to
date it has not been possible to have $C\gg {\bar\lambda}_J$.

For the experiments of Ref.\thinspace\onlinecite{Roberto}
non-symmetric annular Nb/Al-AlO$_{x}$/Nb JTJs were fabricated with
a whole-wafer process in which the junctions are formed in the
window of the SiO insulating layer between the base and the
counter electrodes. See Ref.\onlinecite{Roberto} for details. The
ring-shaped junctions had a mean radius ${\bar r}\cong
80\thinspace\mu$m, width $\Delta r=4\thinspace\mu$m and a geometry
in which both the base and top electrode have a hole concentric to
the ring. The junction temperature was raised above its critical
value ($T_{c}\simeq 9.2\thinspace K$) by means of a heating
resistor placed close to the (unbiased) sample; then the
temperature was lowered by letting the sample cool down by
exchanging heat with the liquid helium bath through some helium
gas. By changing the helium gas pressure it was possible to vary
the thermal constant and hence the quench time $\tau _{Q}$ in the range $%
10^{-2}$s-1s. The samples were measured in a very well
electrically and magnetically shielded environment.  The number of
fluxons trapped during the transition was determined simply by
feeding a proper current to the junction and measuring its
voltage.

Two different samples were produced, both with $1-T_{c_{1}}/T_{c_{2}}\simeq
0.02$. For sample B, we estimated $\xi _{0}=6.5\thinspace\mu$m, $\overline{c}%
_{0}=10^{7}\thinspace$m/s and $\tau _{0}=0.65\thinspace$ps. Although $\overline{t}_J\simeq \overline{%
t}_{s}$, if we take ((\ref{lJbar})) at face value, we find $\overline{\lambda }%
_{J}\simeq 1\thinspace$mm (with experimental uncertainty of up to
$50\%$).[ On the
other hand, if we estimate $\overline{\lambda }_{J}$ at $\overline{\lambda }%
_{J}(\overline{t}_{s})$ we find $\overline{\lambda }_{J}\simeq
0.7\thinspace$mm, identical at our level of approximation.]

With $C\sim \overline{\lambda }_{J}$ we cannot use (\ref{Dn}) as it stands.
However, we would expect to see a fluxon several percent of the time. In
practice (invariably single) fluxons formed once every 10-20 times.

In Sample A, having a similar circumference and similar
superconductors coupled differently, whereby $\overline{\lambda
}_{J}$ was 2.5 times larger, there was much less fluxon trapping,
with no reliable fluxons seen in several quenches (enough to have
seen something in the former case). This demonstrates that the
superconductors cannot be treated individually for fluxon
production, as our earlier discussion confirms. We are imprecise
in the quantitative details since the aim was not to count the
frequency with which defects occurred, but just to have defects at
all. Nonetheless, it is apparent that, in order to have
$\overline{\lambda }_{J}\ll C$ for annular JTJs of a size
comparable to the one above, we must find ways to reduce
$\overline{\lambda }_{J}$ further, as well as possibly increase
$C$.

The quench time $\tau _{Q}$ is difficult to establish accurately.
Fortunately, the small power of $\tau _{Q}/{\bar\tau} _{0}$ in
(\ref{lJbar}) makes the uncertainty largely irrelevant. However,
it has the consequence that a huge reduction in the quench time is
necessary to have any observable effect on $\overline{\lambda
}_{J}$. Similarly, we can gain little from increasing the
asymmetry of the JTJ, even if that is easily
possible experimentally. In consequence, the only realistic way to reduce $%
\overline{\lambda }_{J}$ is by reducing ${\bar\xi} _{0}$ of
(\ref{xi}). Primarily, we need to increase $J_{c}(0)$ to the
largest possible value which does not degrade the barrier quality,
$J_{c}\simeq 10^{4}A/$cm$^{2}$. However, for such currents we
shall have ${\bar t}_s > {\bar t}_J$. The simplest assumption is
that it will be ${\bar t}_s$ that will set the time-scale at which
fluxons are formed. New experiments with such values are under
active consideration by us at this time.

We consider this an important test of the Zurek-Kibble programme
for establishing common constraints in the production of defects
in condensed matter physics and relativistic quantum physics,
given that the spontaneous creation of flux has not been seen in
high $T_{c}$ planar superconductors at the level suggested by the
Zurek-Kibble bound\cite {Polturak} (in fact not at all), and that
vortices  are not observed in $^4 He$ quenches. This is even
though a separate experiment\cite{polturak} supports the phase
separation of the Kibble mechanism.

We conclude with one final concern.  Although the absence of
fluxons in our Sample A, with larger $\overline{\lambda }_{J}/C$,
suggests that the observed fluxons are not artefacts of the
heating and cooling environment, we might be concerned at the
effects of temperature inhomogeneities. In general, moving phase
boundaries leave less (or no) defects in their
wake\cite{volovik2}.

The critical slowing down of $\overline{c}(t)$ and $c(t)$ weakens
the effect of any large scale inhomogeneities in temperature, even
though it is modified slightly for idealised non-symmetric JTJs.
It is difficult to put quantitative limits on the permissable
inhomogeneities but, with empirically comparable $\xi _{0}$, $\tau
_{0}$, and $\tau _{Q}$, the situation is no better of worse for
JTJs than for any other superconducting system undergoing a
mechanical quench.  A possible check of these ideas could be
performed by cooling with a temperature gradient across the ring
circumference, although this would cause its own experimental
problems. We postpone such considerations to the far future.

\section*{ACKNOWLEDGEMENTS}

E.K. and R.R. thank the EU Erasmus/Socrates programme for financial support
and the University of Salerno for hospitality. This work is the result of a
network supported by the European Science Foundation.





\end{document}